\documentclass[%
 aip,
 sd,%
 amsmath,amssymb,
 reprint,%
]{revtex4-1}

\usepackage{graphicx}
\usepackage{dcolumn}
\usepackage{bm}
\usepackage{color} 
\usepackage{hyperref}
\begin{document}

\preprint{AIP/123-QED}

\title[]{Maximizing the quality factor to {mode} volume ratio
	 for ultra-small photonic crystal cavities  }

\author{Fengwen Wang}
 \email{fwan@mek.dtu.dk}
 \altaffiliation{Section of Solid Mechanics, Department of Mechanical Engineering, Technical University of Denmark, {2800 Kgs. Lyngby, Denmark} }
\author{Rasmus Elleb{\ae}k Christiansen}%
\affiliation{Section of Solid Mechanics, Department of Mechanical Engineering, Technical University of Denmark, {2800 Kgs. Lyngby, Denmark}
}%
\author{ Yi Yu}
\affiliation{DTU Fotonik, Department of Photonics Engineering, Technical University of Denmark, {2800 Kgs. Lyngby, Denmark}}%

\author{ Jesper M{\o}rk}
\affiliation{DTU Fotonik, Department of Photonics Engineering, Technical University of Denmark, {2800 Kgs. Lyngby, Denmark}}%

\author{Ole Sigmund}
\affiliation{Section of Solid Mechanics, Department of Mechanical Engineering, Technical University of Denmark, {2800 Kgs. Lyngby, Denmark}
}%

\date{\today}

\begin{abstract}
Small manufacturing-tolerant photonic crystal cavities are systematically designed using topology optimization to enhance the ratio between quality factor and mode volume, ${Q/V}$. 
For relaxed manufacturing tolerance,  a cavity with bow-tie shape is obtained  which confines light beyond the diffraction limit  into a deep-subwavelength volume. Imposition of a small manufacturing tolerance still results in efficient designs, however, with  diffraction-limited confinement. Inspired by  numerical results,  an elliptic ring grating cavity concept is extracted via geometric fitting. Numerical evaluations demonstrate that for small sizes,  topology-optimized cavities enhance  the \textit{Q/V}-ratio by up to two orders of magnitude relative to standard $L1$ cavities and more than one order of magnitude relative to   shape-optimized $L1$ cavities. An increase in cavity size can enhance the \textit{Q/V}-ratio by an increase of the \textit{Q}-factor without significant increase of  \textit{V}. Comparison between optimized and reference cavities illustrates that significant reduction of \textit{V}  requires big topological changes in the cavity.

\end{abstract}

\maketitle


 Strong light-matter interaction is   key in  a wide range of photonic and optoelectronic applications, including low threshold lasers  \cite{Painter1999,Ota2013,Matsuo2013,Xue2016}, sensors \cite{Pitruzzello2018}, nonlinear optics~\cite{Soljacic2004},  cavity quantum electrodynamics~\cite{Tiecke2014}, switching~\cite{Husko2009,Yu2015} and optomechanics \cite{Kippenberg2008}. In a cavity, the local photon density of states (LDOS) scales proportionally to the \textit{Q/V}-ratio. An increase of LDOS in a cavity  can lead to enhanced spontaneous emission through the Purcell effect~\cite{Purcell1946}.  Both photonic crystal  (PhC)   and plasmonic cavities have been used to enhance the Purcell effect~\cite{Akahane2003,Dharanipathy2014,Minkov2014,Khurgin2015}.  PhC cavities increase the temporal confinement of light in a material, as represented by their high \textit{Q},  and  are restricted  by  the diffraction-limited spatial confinement of the light, measured in terms of  \textit{V}~\cite{Akahane2003,Dharanipathy2014,Minkov2014}. Plasmonic cavities  are capable of increasing  the spatial confinement  beyond the diffraction limit, i.e. a low \textit{V} can be attained, but are restricted to small \textit{Q}-values due to ohmic losses \cite{Khurgin2015}. Miniaturization of cavities with a high  \textit{Q/V}-ratio is in demand to  enhance the light-matter interaction and reduce footprint for compact integrated optical circuits.

Previously, many studies considered the design of dielectric PhC cavites with enhanced \textit{Q} while keeping a diffraction-limited  $V \sim (\lambda/n)^3$.  Most of the studies focused on shape optimization (SO) by changing locations or radii of air holes~\cite{Akahane2003,Dharanipathy2014,Minkov2014} or using gradient-based geometry projection methods~\cite{Frei2008,Wang2013}. These studies mainly focused  on the conventional $Ln$ or  $Hn$ cavities~\cite{Akahane2003,Dharanipathy2014,Minkov2014,Frei2008,Wang2013}  ($n$ is the number of filled holes in the PhC) and assumed   very large in-plane dimensions, thus large footprint to reduce in-plane loss. Recently,  bow-tie shaped PhC cavities consisting of two tip-to-tip opposite components were studied~\cite{Gondarenko2008,Lu2013,Hu2016,Choi2017}. However the low \textit{V} in these cavities is found to be highly dependent on the features between the two tips and hence extremely sensitive to manufacturing variations.  3D electron beam lithography (EBL) can fabricate PhC structures with hole size  down to  about 40 nm~\cite{Yamazaki2015} and more recently the width of the bow-tie tip connection  has been controlled to 12 nm with an error region of $\pm 5$ nm~\cite{Hu2018}. The fabrication accuracy puts a limit on the resolution of the tip-region, which is important to take into account when optimizing the design.

\begin{figure}
 	\centering
\includegraphics{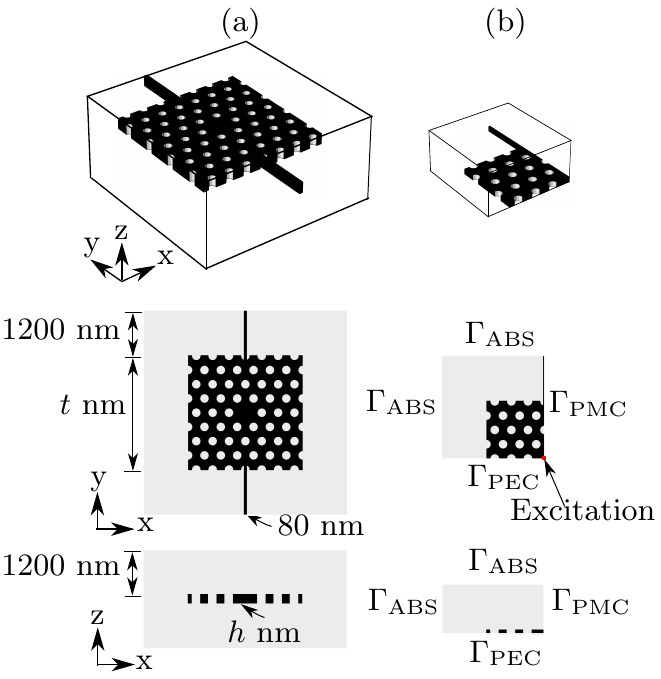}
	\caption{\label{fig:model} Illustration of a 3D PhC membrane and the corresponding computational domain. (a) 3D PhC membrane. (b)~Computational domain and boundary conditions.  }
\end{figure}

A density-based topology optimization (TO) method was also employed to design finite-size PhC cavities with enhanced \textit{Q/V}-ratio~\cite{Liang2013}. Even though these optimized PhC cavities exhibit stronly enhanced performance,  they are  difficult to manufacture, as they contain prohibitively small holes or holes with irregular patterns and sharp features.  Such features, largely determined by the underlying mesh resolution,  may also result in  large modelling errors and erroneous performance estimates.  In this study, we employ TO with manufacturing and length-scale control  to systematically design manufacturable 3D PhC membrane cavities demonstrating an increase of the \textit{Q/V}-ratio by up to two orders of magnitude relative to the standard $L1$ cavity contained within a square membrane
(see Fig.~\ref{fig:model}) and more than one order of magnitude relative to a SO $L1$ cavity.  In addition we investigate the influence of the cavity size on the performance of the optimized and reference cavities.
 
In the limit of a low-loss cavity, the \textit{Q/V}-ratio is only an approximation to the LDOS dominated by the contribution of a single resonant mode~\cite{Purcell1946,Liang2013}. The \textit{Q/V}-ratio also describes the  spontaneous emission enhancement in the case where the emitter linewidth is narrower than  the cavity linewidth~\cite{Mork2018}.  Hence, to directly target the goal of achieving strong light-matter interaction, the design goal is recast as the problem of maximizing a frequency-averaged LDOS in the PhC membrane excited by a dipole~\cite{Liang2013}. Using complex analysis it can be shown that the LDOS may be calculated from the electric field  using a simple multi-pole evaluation instead of a full frequency domain integration~\cite{Liang2013}.  By introducing a squared Lorentzian weight function this evaluation is reduced to a single complex frequency evaluation of the electric field to obtain the LDOS for a given material configuration. The width of the weight function is gradually reduced over the course of the optimization using a continuation scheme to attain the final design for the targeted frequency~\cite{Liang2013}.  

The cavity design and evaluation take offset in the PhC membrane  illustrated in Fig.~\ref{fig:model} with a finite size of $t  \times t  \times h  $. The lattice constant  is set to  $a=440 \ {\rm{nm}}$ and the membrane thickness is set to $h/a= 0.568$.  To avoid a free floating membrane, a thin beam  of width 80 nm is attached to the membrane as shown in the top figure in Fig.~\ref{fig:model}(a). The dielectric material constituting the membrane is Indium Phosphide (InP) with a refractive index of $n_{\rm{InP}}=3.17$. The membrane is surrounded by air ($n_{\rm{air}}=1$) in all  directions. Due to the imposition of symmetry, only one eighth of the full domain is modelled with appropriate boundary conditions as illustrated in Fig.~\ref{fig:model}(b), where $\Gamma_{\rm{PMC}}$ and $\Gamma_{\rm{PEC}}$ denote a  perfect  magnetic conductor  (PMC, $\boldsymbol{n}\times\boldsymbol{H}=\boldsymbol{0}$) and a perfect  electric conductor (PEC, $\boldsymbol{n}\times\boldsymbol{E}=\boldsymbol{0}$), respectively. Absorbing boundary conditions, $\Gamma_{\rm{ABS}}$, are used to truncate the computational domain. The membrane is excited  by a dipole oriented along the x-direction with an angular frequency,  $\omega$, modelled as a current $ \boldsymbol{J}(\boldsymbol{x})  \sim \boldsymbol{e}_x  e^{−{i\omega t}}\delta \left(\boldsymbol{x}-\boldsymbol{x}_0 \right)$. The total electric field is governed by 
\begin{equation} \label{Eq:MaxWell}
  \nabla\times\frac{1}{\mu(\boldsymbol{x})}\nabla\times\boldsymbol{E}(\boldsymbol{x})-{\varepsilon(\boldsymbol{x})}\omega^2\boldsymbol{E}(\boldsymbol{x})=i\omega\boldsymbol{J}(\boldsymbol{x}) \\
\end{equation}


The finite element method is employed to solve  the scattering problem in Eq.~\eqref{Eq:MaxWell}. The solver is implemented using the parallel sparse-matrix library PETSc~\cite{petsc-user-ref} and the MUltifrontal Massively Parallel sparse direct Solver (MUMPS)~\cite{MUMPS:1}
%

Based on the finite element discretization, a continuous design variable, $\rho\in\left[0,\ 1\right]$, is introduced in each element to represent the material occupation in the element, and the elemental refractive index is interpolated using~\cite{Christiansen2018}
\begin{equation}\label{Eq:Inte}
n \left(\rho \right)=\rho \left( n_{\rm{InP}}-  n_{\rm{Air}}\right) +n_{\rm{Air}}
\end{equation}

The design is iteratively updated using the gradient-based optimization method, the Method of Moving Asymptotes (MMA)~\cite{Svanberg1987}. The sensitivities of the objective and constraints are derived using adjoint sensitivity analysis~\cite{Jensen2011,Liang2013}. {The software used to design cavities including the finite element solver and MMA implementation, is developed in-house.} 
  
In order to verify the performance of the optimized cavities,  they  are imported into the commercial software  COMSOL Multiphysics 5.3  and evaluated using eigenvalue analysis~\cite{Lasson2018}. The TO  designs are  extracted from the FEM model by performing an isometric mapping and extracting the design for the isovalue of 0.5 in a post processing step. This avoids stair-casing caused by the FEM mesh of the optimization model which in turn would limit accurate production.  

%
 
\begin{figure}[!htb]
   	 	\centering
   	\includegraphics{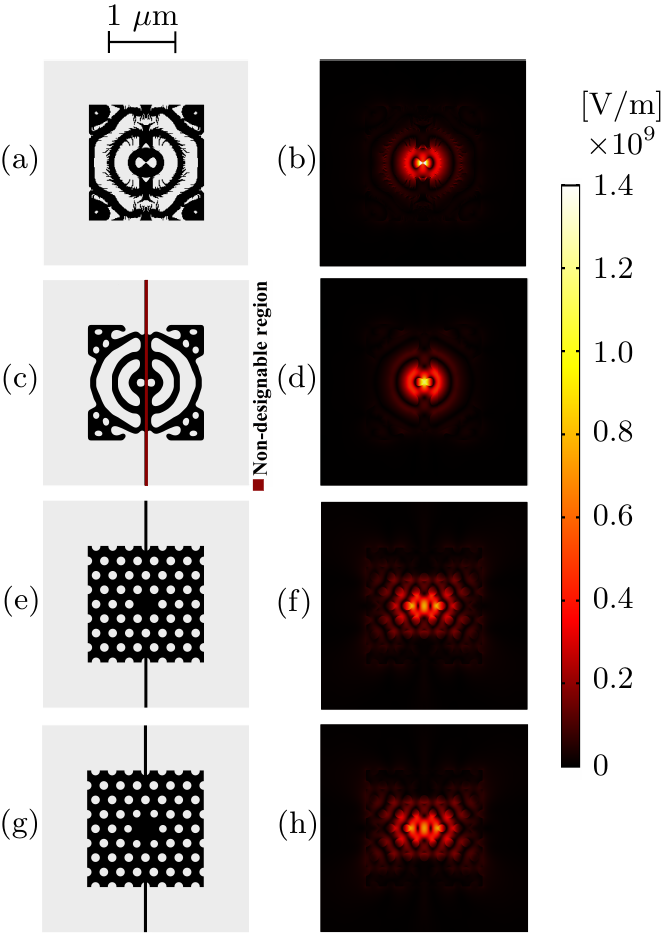}
 	\caption{\label{fig:Opt1540}  Left column: TO cavities and reference ones with $t/a=7$. Right column: Electric field norm $\left\| \boldsymbol{E} \right\| $  of  corresponding  fundamental resonant modes. (a)-(b) TO 3D PhC membrane cavity with small features. (c)-(d) TO 3D PhC membrane cavity with a  minimum length scale of 74 nm. (e)-(f) $L1$ cavity. (g)-(h) SO $L1$ cavity. }
 \end{figure}

As the first optimization case,  we  explore the possible performance attained by TO cavities. Hence we design a free floating PhC cavity  with a size of  $t/a=7$ allowing  small and sharp features. The target wavelength is $\lambda=1550\ \rm{nm}$. As the initial guess for the optimization, we use the  reference $L1$ cavity shown in Fig.~\ref{fig:model}  consisting of a hexagonal lattice with an air cylinder radius of  $r=0.26 a$  without the beam in the center region. The full PhC region is freely designable. The center of the optimized cavity shown in Fig.~\ref{fig:Opt1540}(a)  resembles a bow-tie  with a tip gap of 10 nm at the cavity center, which is  surrounded by a ring-like grating. As expected, small  features are observed in the  optimized cavity, which are likely impossible to fabricate accurately using present day techniques. 

 To avoid the small features present in the previous design,  a minimum length scale is introduced in the design by applying a smoothened Heaviside projection filtering technique to the design field~\cite{Wang2011}.  The smallest feature that can be fabricated is determined mainly by  the EBL and the semiconductor dry etching. As for our InP platform, we found that it is preferable to keep this value larger than 50 nm, otherwise, it will be difficult to make it etched through. Here, a filter  radius of 96 nm is employed,  and a minimum length scale of 74 nm is enforced in both dielectric material and air using a geometrical constraint approach~\cite{Zhou2015}. 
 Further, an 80 nm wide non-designable region occupied by InP  is introduced at the center of the cavity,  highlighted in the red in the optimized design in Fig.~\ref{fig:Opt1540}(c), as the mode should be confined inside the solid and not the air region. This has the effect that the size of the bow-tie region is changed. 
 It is seen that all the features in the design are smooth and conform to the imposed length scale.  
 The  smoothened bow-tie is surrounded by a smooth elliptic ring grating as well as (less important) corner PhC-like regions. 

To evaluate the performance of the optimized cavities, we use the standard $L1$ cavity shown in Fig.~\ref{fig:model}(a) (also shown in  Fig.~\ref{fig:Opt1540}(e)) as a reference ($L1$ has an antinode of $\left\| \boldsymbol{E} \right\|$  at the cavity center while $H0$ has an node~\cite{Minkov2014}).  As an additional reference we use a SO $L1$ cavity obtained using a parameter sweep  over the hole radius ($r_0$) and lattice constant ($a_0$) of the air cylinders closest to the cavity center, with $r_0\in  \left[0.22 a, 0.32a\right] $  and $a_0\in  \left[ 220, 500\right] \ \rm{nm}$. The  SO $L1$  cavity with largest  \textit{Q/V}-ratio is obtained for   $ r_0=0.23 a$  and $a_0=  460 \ \rm{nm}$ and is shown in Fig.~\ref{fig:Opt1540}(g).

 The normalized electric field norm of the resonant modes in the optimized and reference cavities are shown in the right column of Fig.~\ref{fig:Opt1540}. The  performance of the  cavities is summarized in Table~\ref{tab:1540}.  It is seen in  Fig.~\ref{fig:Opt1540}(b) that the cavity with small features  exhibits an extremely high field intensity at its center.  Hence the optimized cavity can concentrate  light into a deep-subwavelength volume with  $V=  0.00026   \left(  {\lambda}/{n}\right)^3$.  Here  $n$ is  the refractive index of the cavity material.  The mode volume is calculated using~\cite{Hu2016} $V= \frac{\int\varepsilon(\boldsymbol{x}) \left|\boldsymbol{E}(\boldsymbol{x})\right|^2 d \boldsymbol{x}}{ \max\left\{\varepsilon(\boldsymbol{x}) \left|\boldsymbol{E}(\boldsymbol{x})\right|^2\right\}} $. 
 Moreover  the surrounding grating structure adapted to the bow-tie shape further reduces field intensity away from the cavity center, which leads to an increase in \textit{Q} and a further reduced \textit{V} for the targeted resonant mode. Hence the optimized cavity shown in Fig.~\ref{fig:Opt1540}(a) displays at least two order  of magnitude lower \textit{V} than the optimized $H0$ cavity by Wang et  al.~\cite{Wang2013} and five time smaller \textit{V} than the other proposed cavities by Gondarenko and Lipson~\cite{Gondarenko2008}. Further, it possesses a high \textit{Q/V}-ratio, which is  at least two times that of the highest \textit{Q/V}-ratio obtained for a dielectric bow-tie cavity proposed by Lu et al.~\cite{Lu2013}  and a hybrid photonic-plasmonic nanobeam cavity at room temperature by Conteduca et al~\cite{Conteduca2017}. It is known that \textit{V} is mainly determined by the size of the  gap in the bow-tie shape at the cavity center. The smaller the gap, the smaller \textit{V}~\cite{Lu2013}. In this work, the numerical resolution and post processing limited the gap size between two bow-tie tips to 10 nm. 
 
The field intensity at the cavity center for the design with restricted spatial resolution, shown  in Fig.~\ref{fig:Opt1540}(c)-(d), is lower than for the previous design shown  in Fig.~\ref{fig:Opt1540}(a)-(b). However, it is still significantly larger  than the $L1$ and  SO $L1$ cavities shown in  Fig.~\ref{fig:Opt1540}(e)-(h). Moreover the spatial confinement of the field pattern is smaller than in the reference cavities. Hence,  it is evident that the optimized cavity possesses a significantly enhanced \textit{Q/V}-ratio  relative to both reference cavities (see Table~\ref{tab:1540}). However, the introduction of the non-design domain at the  cavity center and   the enforcement of length scale  prohibits the appearance of small features, which in turn restricts  the optimized design from attaining a \textit{V} below the diffraction limit of $V\sim 0.1 \left(  {\lambda}/{n}\right)^3$. It is observed that  the TO cavity with small features confines light in the air regions while  the one with imposed length scale confines light inside the InP. The non-design domain guarantees the confinement of light in the material region at the  cavity center, which is important for applications~\cite{Matsuo2013}.

 
 Compared to the standard $L1$ cavity, the SO $L1$ cavity enhances the \textit{Q/V}-ratio by an increased  \textit{Q} with a small increase in \textit{V}. The TO cavity with imposed length scale enhances the \textit{Q/V}-ratio by both   increased \textit{Q} and   reduced \textit{V} and displays  a \textit{Q/V}-ratio at least 18 times higher than both  references. The finite size of the PhC membrane  limits the \textit{Q}-value  due to in-plane loss. The elliptic  Bragg gratings created in the TO cavities reduces in-plane loss hereby facilitating a significant increase in \textit{Q}.

\begin{table}
\caption{\label{tab:1540}Performance of the TO, $L1$ and SO $L1$ cavities with $t/a=7$ for the fundamental resonant mode }
\begin{ruledtabular}
\begin{tabular}{l c c c c}
 Design&$\lambda$  [nm] &\textit{Q} &\textit{V} $ \left[  \left(  {\lambda}/{n}\right)^3 \right] $ & $Q/V$  $\left[\left( {n}/{\lambda}\right)^3\right]  $ \\
\hline
TO (a) &   1549&   1062   &   0.0002651 &    $ 4006\times10^3$\\
TO  Len. (c) & 1554 &  2979 &    0.1083 &   $27.50\times10^3$    \\
$L1$ (e)&     1451  &  265.4   &   0.3142 &     $0.8446\times10^3$ \\
SO $L1$ (g) &1506  &   486.7  &    0.3330 &    $1.462\times10^3$ \\
\end{tabular}
\end{ruledtabular}
\end{table}
Next we study the effect of the outer dimensions of the PhC cavity on the attainable \textit{Q/V}-ratio. 

\begin{figure}
 	 	\centering
 	\includegraphics{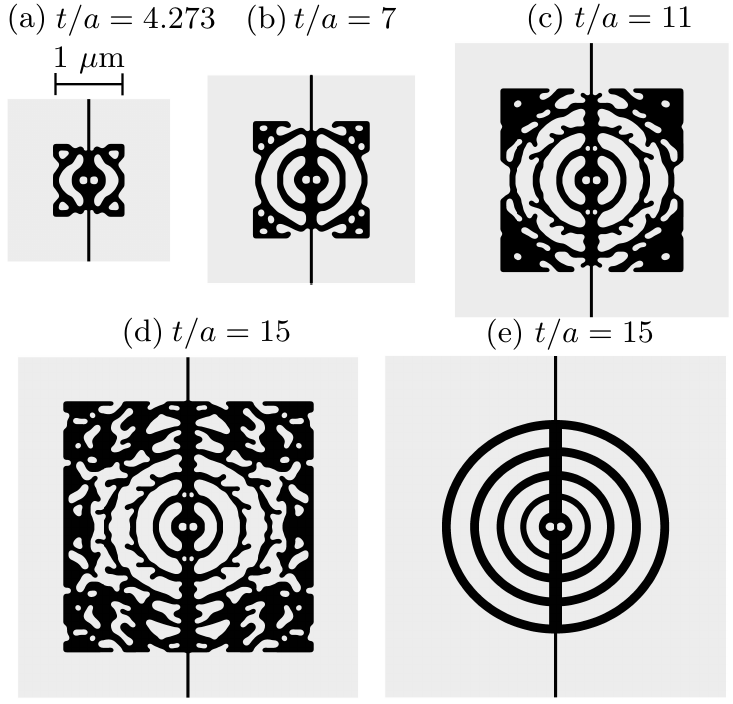}
	\caption{\label{fig:TopOpt}  (a)-(d) TO 3D PhC membrane cavities with different sizes. (e) Extracted elliptic ring grating cavity.}
\end{figure}

TO cavities, designed with resolution limitation  and with different maximum outer dimensions of  $t/a=4.273$,  $t/a=7$, $t/a=11$  and $t/a=15$ are presented in Fig.~\ref{fig:TopOpt}(a)-(d). It is seen that all the optimized cavities retain bow-tie shape region surrounded by the  elliptic ring gratings adapted to a central bow-tie shape, indicating the importance of both types of features in attaining a high \textit{Q/V}-ratio. More complex geometric features are seen to appear in the largest cavity (Fig.~\ref{fig:TopOpt}(d)). These intricate features are however less important for the design performance due to the low field intensity further from the cavity center.

By extracting the elliptic rings and bow-tie feature from the TO cavity in Fig.~\ref{fig:TopOpt}(d) and fitting a simplified elliptic grating structure with varying bar width and spacing,  an elliptic ring grating cavity is obtained, see~Fig.~\ref{fig:TopOpt}(e). Corresponding  elliptic ring grating cavities for $t/a=11 $, $t/a=7 $, $t/a=4.273 $ are obtained by removing one, two and three elliptic rings from the design in Fig.~\ref{fig:TopOpt}(e). 

\begin{figure}
 	 	\centering
 	\includegraphics{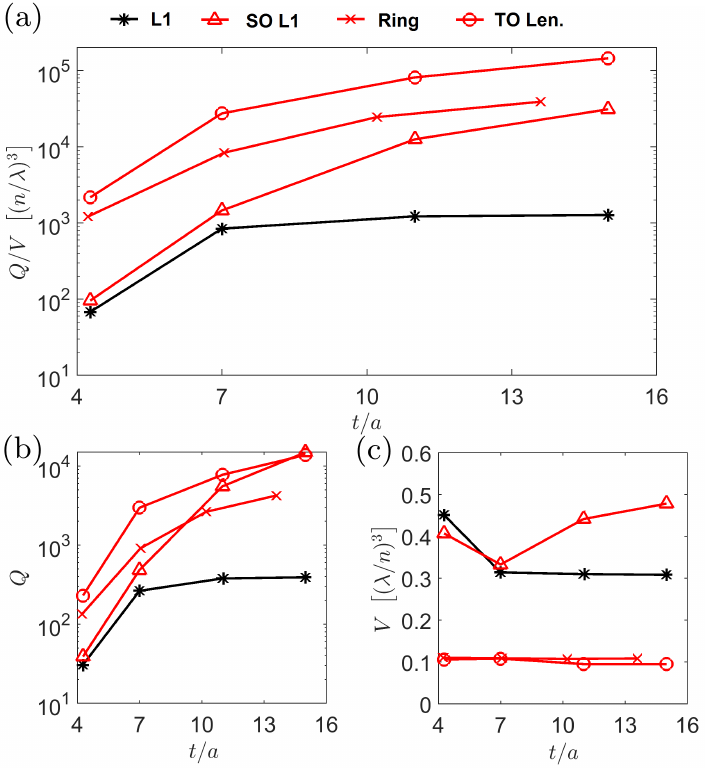}
 	\caption{\label{fig:Summary}  Overall performance comparison of different cavities vs cavity size. (a) \textit{Q/V} vs cavity size. (b) \textit{Q} vs cavity size. (c) \textit{V} vs cavity size.}
 \end{figure}

The performance of the reference, TO with resolution limitation  and parametrized ring cavities are shown in Fig.~\ref{fig:Summary}. SO $L1$ cavities are obtained by performing a SO for each individual size. {Note that the sizes of the elliptic ring grating cavities deviate from the other cavities. The \textit{t/a}-values reported in Fig.~\ref{fig:Summary} correspond to the dimension in the $x$-direction.}  As expected, the larger the cavity size, the larger the \textit{Q/V}-ratio. Unsurprisingly, the TO cavities are seen to  exhibit the best performance followed by the elliptic ring grating cavities.  The elliptic ring grating cavities possess smaller \textit{Q} compared to the corresponding TO cavities, but have similar \textit{V}. Moreover, it is seen that increasing cavity size can enhance \textit{Q} significantly, however, \textit{V} does not change much since the main contribution to the low \textit{V} stems from the bow-tie at the cavity center. For the largest size, there is no significant  difference  in terms  of \textit{Q} between the TO and SO cavities since the in-plane loss is very small in the large cavity and the \textit{Q} is mainly dominated by out-of-plane loss. Comparing  the TO designs  to the $L1$ and  SO $L1$ cavities, a significant decrease in \textit{V} is observed  for all sizes.  This indicates that to attain a lower \textit{V}, significant geometric changes to the cavity are required.  Moreover the extracted elliptic ring grating cavities exhibit  more in-plane losses represented by smaller  \textit{Q} than the SO $L1$ and TO cavities for large cavity size. This implies that the smaller branches attached in the ring gratings and the features at the membrane corners  in Fig.~\ref{fig:TopOpt} (c)-(d) can facilitate a higher \textit{Q}. However, the simplicity of the parametrized elliptic ring cavity may be appealing from a fabrication point of view. 
 
In conclusion, a density-based TO method has been employed to the  systematic  design of manufacturable small  finite-size membrane cavities  with a smallest feature size of 74 nm, well above standard EBL tolerances~\cite{Yamazaki2015}. Further, the influence of cavity size on \textit{Q} and \textit{V} was investigated. Inspired by the TO designs,  elliptic ring grating cavities  were extracted using parametric models.  The overall comparison of cavity performances demonstrates that the TO cavities perform best among all the cavities considered  followed by  the extracted ring grating cavities. By increasing the cavity size, one can obtain higher \textit{Q}, however there is no significant improvement in \textit{V}. The lower \textit{V} requires significant geometric  changes in the cavity center, such as from $L1$ to elliptic ring grating cavities.  Moreover resolution restrictions lead to near-diffraction-limited volumes. Smaller features are required to reach a deep-subwavelength mode volume.  By reducing the minimum length  scale and introducing a smaller fixed material region at the cavity center it would be possible to obtain designs with significantly lower \textit{V} while ensuring that the field is confined to the solid.  {Even though this study is mainly focused on designing cavities with field confinement in solid, the whole procedure can be used to design cavities with field confinement in the air as well, suitable for other applications such as optical tweezers~\cite{Juan2011}.} 

This work was financially supported by Villum Fonden via the NATEC  Center of Excellence (grant 8692).
%
\end{document}